\documentclass[conference]{IEEEtran}
\IEEEoverridecommandlockouts
\usepackage{cite}
\usepackage{amsmath,amssymb,amsfonts}
\usepackage{graphicx}
\usepackage{textcomp}
\usepackage{xcolor}
\def\BibTeX{{\rm B\kern-.05em{\sc i\kern-.025em b}\kern-.08em
    T\kern-.1667em\lower.7ex\hbox{E}\kern-.125emX}}

\usepackage[linesnumbered,lined,ruled,commentsnumbered]{algorithm2e} 

\newtheorem{definition}{Definition}
\usepackage[numbers]{natbib}

\begin{document}

\title{Spatially and Robustly Hybrid Mixture Regression Model for Inference of Spatial Dependence 
}

\author{\IEEEauthorblockN{Wennan Chang$^1$, Pengdao Dang$^1$, Changlin Wan$^1$,
Xiaoyu Lu$^2$, Yue Fang$^3$, \\ 
Tong Zhao$^4$, Yong Zang$^3$,  Bo Li$^5$, Chi Zhang$^1$ and Sha Cao$^3$
}
\IEEEauthorblockA{
$^1$Electrical and Computer Engineering, Purdue University
$^2$School of Informatics and Computing, Indiana University\\
$^3$Department of Biostatistics, Indiana University, Indianapolis, USA, 46202\\
$^4$Amazon, Seattle, USA 
$^5$School of Economics, Peking University, Beijing, China\\
Corresponding Author Email: czhang87@iu.edu, shacao@iu.edu
}}

\maketitle
\thispagestyle{plain}
\pagestyle{plain}

\begin{abstract}

In this paper, we propose a Spatial Robust Mixture Regression model to investigate the relationship between a response variable and a set of explanatory variables over the spatial domain, assuming that the relationships may exhibit complex spatially dynamic patterns that cannot be captured by constant regression coefficients. Our method integrates the robust finite mixture Gaussian regression model with spatial constraints, to simultaneously handle the spatial nonstationarity, local homogeneity, and outlier contaminations. Compared with existing spatial regression models, our proposed model assumes the existence a few distinct regression models that are estimated based on observations that exhibit similar response–predictor relationships. As such, the proposed model not only accounts for nonstationarity in the spatial trend, but also clusters observations into a few distinct and homogenous groups. This provides an advantage on interpretation with a few stationary sub-processes identified that capture the predominant relationships between response and predictor variables. Moreover, the proposed method incorporates robust procedures to handle contaminations from both regression outliers and spatial outliers. By doing so, we robustly segment the spatial domain into distinct local regions with similar regression coefficients, and sporadic locations that are purely outliers. Rigorous statistical hypothesis testing procedure has been designed to test the significance of such segmentation. Experimental results on many synthetic and real-world datasets demonstrate the robustness, accuracy, and effectiveness of our proposed method, compared with other robust finite mixture regression, spatial regression and spatial segmentation methods.

\end{abstract}


\section{Introduction}
Many problems in the environmental, economic, and biological sciences involve spatially collected data, and a main problem of interest is investigation of the relationship between a response variable and a set of explanatory variables over the spatial domain using regression modeling. Notably, the relationships between response variables and covariates may exhibit complex spatially dynamic patterns that cannot be captured by constant regression coefficients. Instead, such relationships may abruptly change at a certain boundary of two neighboring spatial clusters, but stay relatively homogeneous within clusters. Detecting clusters of observations that display similarity in both regression relationships and spatial proximity allows straightforward interpretations of local associations between response variables and covariates. For example, the residential real estate pricing could be quite similar in a local community, but drastically differ for two houses across the street \cite{dolde1997temporal}; a major goal of analyzing functional magnetic resonance imaging (fMRI) data is to detect spatially distributed and functionally linked regions that continuously share information with each other in reaction to different stimuli \cite{van2010exploring}. For all these real-world application settings, the collected data may often contain outliers, which may severely corrupt the analysis results if not properly handled. Overall, the spatial nonstationarity, local homogeneity, and model robustness are three main challenges in spatial regression modeling.

In the nonspatial setting, finite mixture regression models have been used in many areas as an effective exploratory approach to identify heterogeneity in response–predictor relationships. For an overview, see \cite{mclachlan2004finite,fruhwirth2006finite}. To account for outliers or heavy-tailed noises, many algorithms have been developed to estimate the parameters robustly \cite{yu2020selective}. To seek for robust parameter estimation in the presence of outliers, methods have been developed that replaced the least-square criterion in the M-step of the expectation maximization (EM) algorithm by more robust criterion \cite{markatou2000mixture,shen2004outlier,bai2012robust,bashir2012robust,song2014robust,yao2014robust,peel2000robust}. To enable simultaneous model estimation and outlier removal, penalized mean-shift mixture model \cite{yu2017new}, and the least trimmed likelihood estimator \cite{neykov2007robust,dougru2018robust,garcia2010robust, garcia2017robust} were proposed. While these methods could robustly capture the heterogeneous relationship between response and predictor variables, they are not designed to model the spatial dependency. 

In modeling the spatial dependency, conventional nonstationary spatial regression models such as geographically weighted regression (GWR)
\cite{brunsdon1996geographically,fotheringham2003geographically,wheeler2010geographically} and Bayesian spatially varying coefficient (SVC) \cite{fuentes2002spectral,banerjee2014hierarchical} models fit as many regression models to the data as there are observations, at the cost of a large computational burden for large spatial datasets, and sometimes may lead to overfitting. In addition, interpretation of the GWR and SVC models require visual inspection of the coefficient maps to pursue local homogeneity, and can not automatically capture the spatially clustered patterns. In order to automatically detect spatially homogeneity cluster, a penalized spatial regression model has been proposed \cite{li2019spatial}, where a fused-lasso \cite{tibshirani2005sparsity} type of penalty has been developed to account for the spatial homogeneity in the linear regression setting.  Nevertheless, the spatial smoothness assumption in the above spatial regression models could be problematic and violated  due to natural or man-made discontinuities in the spatial domain. In addition, none of these methods is designed to handle outliers. 

Model-based spatial segmentation is another type of methods to deal with spatial data using spatially constrained Gaussian
mixture model \cite{nguyen2011gaussian,nguyen2012fast}. Spatial segmentation incorporates spatial information between neighboring
pixels into the Gaussian mixture model based on Markov random field (MRF), with a goal to cluster all variables (e.g. pixels in image), where the distance of two instances is dependent on both their feature expressions and spatial proximity. This comes at a high computational cost. While robust spatial segmentation algorithms are available \cite{nguyen2011gaussian,nguyen2012fast}, they fail to intentionally model the linear relationship between the response and predictors, but instead simply treat the response and predictors as different features.

In summary, none of the existing methods could robustly model the spatial clustering patterns of linear dependency between response and predictors, and
we propose a novel Spatial Robust Mixture Regression (SRMR) model that enables a simultaneous detection of spatial regions in which variables have a strong linear dependency.

The key contributions of work include: (1) We developed the very first computational concept of spatially dependent mixture regression analysis. (2) We provided the SRMR model that efficiently solves the  spatially dependent mixture regression problem, which is also empowered by a statistical inference approach to assess regression significance.
(3) SRMR enables a new type of spatial segmentation analysis to detect overlapped spatial regions of varied dependencies among subset of features, which have high contextual meaningfulness.

\section{Preliminary}
\subsection{Notations}
We denote scalar value, vector, and matrix as lowercase character \(x\), bold lowercase character \(\textbf{x}\), and uppercase character \(X\), respectively. Let $\{(\boldsymbol{x}(s_{i}), y(s_{i})), i=1, \ldots, n\}$ represent a set of spatial data that is observed at spatial locations $s_{1}, \ldots, s_{n} \in \mathbb{R}^{2}$, where the response variable $y(s_{i})$ is assumed to be spatially correlated,  $\boldsymbol{x}(s_{i})=$ $(x_{1}(s_{i}), \ldots, x_{p}(s_{i}))^{\mathrm{T}}$ is the $p$-dimensional vector of explanatory variables for the observation located at $s_{i}$, and $s_{i}=(c_i^1,c_i^2)$ is the 2-dimensional coordinate of the $i$th location. In this study, we only describe and validate the SRMR model on 2-dimensional spatial data. Noted, the approach can be directly applied to $K$-dimensional ($K>2$) spatial data.

\subsection{Problem statement}

To capture the spatially dependent structure for the response variable, we write a standard generalized linear regression model (GLM) for the $i$-th spatial location as follows,
$$g(E(y=y(s_{i}) \mid \boldsymbol{x}=\boldsymbol{x}(s_{i})))=\sum_{j=1}^{p} x_{j}(s_{i}) \beta_{ji}+\epsilon_i$$
where $\beta_{ji}, j=1,2, \ldots, p$, are the regression coefficients for the $p$ predictors, and $\epsilon_i$ represents random noise with mean 0 and variance $\sigma_{i}^{2}$, and $g(\cdot)$ is the link function. In this work, we assume identify link for linear regression. The intercept can be accommodated by including 1 as an entry of $\boldsymbol{x}(\boldsymbol{s}_{\boldsymbol{i}})$. Apparently, unless with sufficient number of repeated measurements for each location, the $\beta_{ji}, \sigma_i$ are non-identifiable. In many cases, there is only a single observation for each spatial location, certain spatial constraints will be enforced  to ensure the identifiablity of the model parameters.

\begin{definition}\textbf{Spatially Dependent Mixture Regression.} Given a dataset consisting of $n$ observations $\{(\boldsymbol{x}(s_{i}), y(s_{i})), i=1, \ldots, n\}$ from spatial locations $s_{1}, \ldots, s_{n}$, the goal of spatially dependent mixture regression is to identify spatial regions $\Pi_{1}, \ldots, \Pi_{K}$ and the number $K$, s.t.,
$$y(s_{i})=\sum_{j=1}^{p} x_{j}(s_{i}) \beta_{j}^{k}+\epsilon_i,\ if\ s_{i}\in\Pi_{k}$$
, where $\beta_{j}^{k}, j=1,...,p, k=1,...,K$ are regression parameters for the $p$ predictors in the $k$-th cluster; $\epsilon_i \sim \mathcal{N}(0,\sigma_k) $, where $\sigma_k$ represents the noise level of cluster $k$. 
\end{definition}

To account for the presence of outliers, we assume that $\Pi_{K}$ are non-overlapping subsets of the whole set $\{1,...,n\}$, and denote the outlier set as $\Pi_{0}$, such that $\Pi_{0}=\{1,...,n\}\setminus\bigcup_{k=1}^{K} \Pi_{k}$ Two type of outliers will be considered here: \\
\textbf{Type 1 Outliers:} $y(s_{i})\not=\sum_{j=1}^{p} x_{k}(s_{i}) \beta_{j}^{k}+\epsilon^k, \forall k=1,...,K$\\
\textbf{Type 2 Outliers:} $\exists k, y(s_{i})=\sum_{j=1}^{p} x_{k}(s_{i}) \beta_{j}^{k}+\epsilon^k, s_{i}\not\in\Pi_{k}$\\
Here the \textbf{Type 1 Outliers} represent the samples do not fit any regression model while the \textbf{Type 2 Outliers} represent the ones fit a certain model but do not locate  nearby the spatial region.

Noted, pre-assumptions of the spatial regions $\Pi_{1}, \ldots, \Pi_{K}$ are needed to enable a valid solution of the spatially dependent mixture regression problem. Such assumptions include a connected spatial region, a compact shape, or high enrichment to a certain region. Noted, as spatially dependent mixture regression assigns each sample into one spatial region $\Pi_{k}$, it directly forms a spatial segmentation method.

\subsection{Related works}
\textbf{Mixture regression and robust estimators}.  Consider an finite mixture Gaussian regression model parameterized by $\boldsymbol\theta = \{ (\pi_k, \boldsymbol{\beta}_k, \sigma_k^2) \}_{k=1}^K$, the conditional density of $y$ given $\boldsymbol{x}$ is $f(y| \boldsymbol{x,\theta}) = \sum_{k=1}^K \pi_k \mathcal{N}(y; \boldsymbol{x^T \beta_k}, \sigma_k^2)$, where $\mathcal{N}(y; \mu, \sigma^2 )$ is the normal density function with mean $\mu$ and variance $\sigma^2$. Many algorithms have been developed to estimate the parameters robustly \cite{yu2020selective} by replacing the least-square criterion in the M-step with more robust criterion in the EM algorithm \cite{bai2012robust,song2014robust,yao2014robust,peel2000robust}. To enable simultaneous model estimation and outlier detection, usually a hyperparameter regarding the proportion of outlying samples needs to be specified, such as in the penalized mean-shift mixture model \cite{yu2017new}, and the least trimmed likelihood estimator \cite{neykov2007robust, chang2020robmixreg}. 

\textbf{Spatially smooth regression.} Conventional nonstationary spatial regression models such as geographically weighted regression (GWR)
\cite{brunsdon1996geographically,fotheringham2003geographically,wheeler2010geographically} and Bayesian spatially varying coefficient (SVC) \cite{fuentes2002spectral,banerjee2014hierarchical}
models allow regression coefficients to vary smoothly as a function of the spatial domain. For GWR, assuming a linear model with $\boldsymbol y$ denoting the observed response vector and $\boldsymbol{X}$ the design matrix, the regression coefficient at the $i$ th location is estimated from $\widehat{\boldsymbol{\beta}}_{i}=(\boldsymbol{X}^{\mathrm{T}} \boldsymbol{W}_{i} \boldsymbol{X})^{-1} \boldsymbol{X}^{\mathrm{T}} \boldsymbol{W}_{i} \boldsymbol{y}$, where $\boldsymbol{W}_{i}$ is a diagonal weight matrix defined by a kernel function of distance of all other points to point $i$. The challenge with GWR and SVC models is that they
fit as many regression models to the data as there are observations, at the cost of a large computational burden, possible over-fitting and interpretation. A penalized spatial regression model has been developed to automatically detect clusters \cite{li2019spatial} by incorporating a fused-lasso penalty constructed based on spatial proximity.

\textbf{Spatial segmentation.} Model-based spatial segmentation aims perform a segmentation task to all samples (e.g. pixels in image) based on the input features. Model-based spatial segmentation adopts an energy function $U(\Pi)$ to integrate the spatial information such as neighborhoods with a regular clustering analysis of the features. Intrinsically, such methods leverage spatial and data consistency to segment spatial regions, i.e., only considering the covariance of independent variables, which cannot solve the spatially dependent regression problem.

\section{Method}
To solve the problem of spatially dependent mixture regression, computational challenges arise from three aspects: (1) the mixture regression model and spatial consistency do not form one unified likelihood function, which prohibits a direct solution by using EM algorithm, (2) detection spatial regions should depend on both goodness of fitting and spatial consistency, and (3) there is lack of a validate approach to assess the statistical significance of mixture regression models.

\begin{algorithm*}
\caption{Hybrid Mixture Regression (HMR)}\label{hybrid}
\SetKwInput{kwInit}{Initialization}
\DontPrintSemicolon
\KwIn{Response vector ${Y}$; independent variables in matrix $X^{N\times(P+1)}$; the number of mixing component $K$; size of initialization random sample $n_0$; 2-dimentional spatial coordinates $S^{N \times 2}$; hyperparameter $\lambda$. }
\KwOut{Partition $\mathcal{C^*}=\bigcup_{k=1}^K C_k$; Mixture regression model parameter estimate $\boldsymbol\theta^* $; spatial centriod parameter $w^*$ ; type 2 outlier set $U^*$}
\kwInit{$\mathcal{C}=\{C_1,...,C_K\}, C_i \subseteq \{ 1,...,N \}$ based on coordinate $S$; compute centroid point $w=\{ w_1,...,w_K \}$ with $\mathcal{C}$ }

\For{$m = 0,..., L_0$ or until convergence}{
     
  \textbf{E-step}: Compute for $i=1,...,N$ and $k=1,...,K$, the hybrid posterior probabilities $p_{ik}^{(m)}$ by \; $p_{ik}^{(m)} = (1-\lambda)p_{reg}(z_i=k | \boldsymbol{x_i}, y_i, \boldsymbol{\theta}^{(m)}) + \lambda p_{spa}(z_i=k | S_{i,:},w^{(m)})$  \;
  
  \textbf{C-step}: For $i=1,...,K$, assign $C_k^{(m)}= \{ i| \underset{l \in \{ 1,...,K \}}{\operatorname{argmax}} \ p_{il}^{(m)}=k, i=1,...,N\}$,  
  and let $n_k^{(m)}$ be the size of $C_k^{(m)}$, $U_k^{(m)} = \{ i| z_{ir} \neq z_{is}, z_{ir}= \underset{z_{ir} \in \{ 1,...,K \}}{\operatorname{argmax}} p_{reg} , z_{is}= \underset{z_{is} \in \{ 1,...,K \}}{\operatorname{argmax}} p_{spa} \}$  \;
  
  \textbf{M-step}: For $k=1,...,K$, the parameters are then updated by $\pi_k^{(m+1)} = \frac{n_k^{(m)}}{\sum_{l=1}^K  n_l^{(m)} }$,  $( \boldsymbol{\beta}_k^{(m+1)}, \sigma_k^{2(m+1)}) =   \textbf{OLS} (  {{Y}_{C_k^{(m)}}, X_{C_k^{(m)},:}} )$ , $w_k = \frac{1}{n_k^{(m)}} \sum_{i \in C_k} S_{i,:}$
}
\end{algorithm*}

\subsection{SRMR algorithm and mathematical considerations}
In sight of the challenge, we developed the spatial robust mixture regression (SRMR) algorithm to conduct simultaneous outlier detection and spatially dependent mixture regression estimation. The underlying idea is that by assuming a likelihood function of spatial regions $p_{spa}$ and introducing a tuning parameter $\lambda\in(0,1)$ to link $p_{spa}$ with the likelihood of mixture regression $p_{reg}$, a surrogate likelihood function $(1-\lambda)p_{reg}+\lambda p_{spa}$ is developed to enable a modified EM-algorithm (\textbf{Algorithm 1}). The inputs of \textbf{Algorithm 1} include the response and independent variables, spatial coordinates, and the hyper parameter $\lambda$. It conducts a simplified spatially dependent mixture regression fitting by assuming there is only Type 2 outliers, i.e., the sample fit one mixture model but do not locate in the corresponding spatial region. Hence, \textbf{Algorithm 1} fits a conventional mixture regression model and computes the spatial regions that are top enriched by the samples fit each regression component. In this study, we assume the spatial likelihood follows $p_{spa}(z_i=k | s_i,w)\propto ||s_i,w||_2$, where $z_i$ represents the class of sample $i$ and $||s_i,w||_2$ represents the Euclidean distance between the spatial coordinate of the sample $s_i$ and the centers of the spatial regions $w$, i.e., assuming the spatial regions form a compact shape. Specifically, a voting step (C-step) is introduced in \textbf{Algorithm 1}, which identifies Type 2 outliers by the ones whose most likely regression component and spatial region are not consistent. Noted, as all the input samples are utilized in the estimation of the mixture regression model, \textbf{Algorithm 1} is always convergent.

Based on the \textbf{Algorithm 1}, we developed the SRMR framework (\textbf{Algorithm 2}). In SRMR, we iteratively conduct the Type 2 outlier only spatially dependent mixture regression by using the \textbf{Algorithm 1} and identify Type 1 outliers by running a robust linear regression on all the samples predicted to each spatial region. The underlying consideration is that only one regression component is consisted within each identified spatial region, which could be effectively identified by a conventional robust regression approach (\textbf{RLM}). In SRMS, we implement the trimmed likelihood estimation based robust mixture regression. The inputs of SRMR is the same as the input of \textbf{Algorithm 1} plus the maximal iteration number $L_0$ and a random seed. The outputs of SRMR include the identified mixture regression models and outliers. The component of each non-outlier samples can be further assigned by maximal likelihood. In SRMR, we utilize the same BIC function for conventional robust mixture regression analysis.

\subsection{Statistical Inference}
\subsubsection{Hypothesis testing for spatial regions}
We conducted a geometry based approach to estimate the significance to observe a spatial region of a certain size. Noted, we utilized the compact spatial shape assumption in SRMR, which could be considered as a round shape. For a round shape with a diameter of $r$, the number of the shapes needed to cover a rectangular spatial region can be computed by $0.28m\times n / r^2$, which serves as a weight to correct the p value assessed from each single component robust regression as detailed following.

\subsubsection{Hypothesis testing for robust linear regression}
We discuss hypothesis testing of the significance a robust linear regression model parameterized by $\hat{ \boldsymbol{\theta}}=\{\hat{ \boldsymbol{\beta}}, \hat\sigma, \hat{\boldsymbol \eta}\}$, which represents the robust regression coefficients estimator, standard deviation estimator, and the index of the outlying samples respectively. A bootstrap procedure is adopted to test the null hypothesis $\boldsymbol\theta= \hat{ \boldsymbol{\theta}}$. We perform the following steps.

Step 1: Calculating the residuals for all observations, including the outlier samples, under regression parameter $\hat{ \boldsymbol{\beta}}, \hat\sigma$, denoted as $\boldsymbol{\epsilon}=\{\epsilon_1,...,\epsilon_n\}$. Let $\boldsymbol{\epsilon}_{out}$ be the residuals corresponding to outlying samples, and $\epsilon_0$ be smallest absolute residual in $\boldsymbol{\epsilon}_{out}$.

Step 2: Generate iid sample $\tilde{\epsilon}_{1}, \ldots, \tilde{\epsilon}_{n}$ from the normal distribution $\mathcal{N}(0,\hat\sigma)$, denoted as $\tilde{\boldsymbol{\epsilon}}=(\tilde{\epsilon}_{1}, \ldots, \tilde{\epsilon}_{n})$.

Step 3: Calculate the percentage of samples in $\tilde{\boldsymbol{\epsilon}}$ whose absolute values are larger than $\epsilon_0$, and denote it as $p_0$.

Step 4: Repeat steps 2-3 for $B$ times, and the statistical significance is evaluated as the average of $p_0$ for the $B$ times.

\subsection{Discussion}
Several prominent features make our proposed approach attractive. First instead of using a robust estimation criterion or complex heavy-tailed distributions to robustify the mixture regression model, our method is built upon a spatial regression model so as to facilitate computation and model interpretation. Second we adopt a sparse and scale-dependent mean-shift parameterization. Each observation is allowed to have potentially different outlying effects across different regression components, which is very flexible. Compared to existing spatial regression methods, our approach allows an efficient solution via the celebrated penalized regression approach, and different information criteria (such as AIC and BIC) can be used to adaptively determine the proportion of outliers. In the next section, we utilized extensive simulations to demonstrate the performance of SRMR and its highly robustness to both gross outliers and high leverage points.

\begin{algorithm*}
\caption{Spatial Robust Mixture Regression (SRMR)}\label{SRMR}
\SetKwInput{kwInit}{Initialization}
\DontPrintSemicolon
\KwIn{Response vector ${Y}$; independent variables in matrix $X_{N\times(P+1)}$; the number of mixing component, $K$; size of initialization random sample, $n_0$; the maximum number of iteration $L_0$; the number of random starts $J$}
\KwOut{Partition $\mathcal{C^*}=\bigcup_{k=1}^K C_k$; robust FMGR parameter estimate $\boldsymbol\theta^* = \boldsymbol\theta^{J_0}$; outlier set $U^*=U^{J_0}$}

\For{$j = 0,...,J$ }{
    Initialization: $U^{(old)} =\{ 1,...,N \}; U^{(cur)}=\emptyset; L=0$ \;
    \For{k=1,...,K}{
        Draw a random sample of size $n_0$ from set $\{1,...,N\}$, indexed by $I_k$ \;
        Run robust linear regression: $(M_k,\beta_k,\sigma_k)=: \textbf{RLM}(y_{I_K} \sim X_{I_k})$ \;
        Initialize posterior probability: $p_ik=\mathcal{N}(y_i - x_i^T \boldsymbol{\beta}_k; 0, \sigma_k^2 ),i=1,...,N$ \;
    }
    \While{$U^{(old)} \neq U^{(cur)} \& L < L_0 $}{
        Let $U^{(old)}=U^{(cur)};L=L+1$ \;
        \For{k=1,...,K}{
            Let $I_k$ be sample indices most likely in cluster $k$ \;
            Let $U_k^{(cur)}$ be type 1 outliers of $Y_{I_k} \sim X_{I_k,:}$, using least trimmed suares robust regression \;
        }
        $U^{(cur)}_{reg}=\bigcup_k U_k^{(cur)}; S=\{1,...,N\}-U^{(cur)}_{reg}$ \;
        Update $(\boldsymbol{\theta}, w)$ by \textbf{HMR} with the rest of samples in $S$, let $U^{(cur)}_{spa}$ be type 2 outliers, and $W$ be the hybrid posterior probability \;
    }
    $U^j = U^{(cur)}_{reg} + U^{(cur)}_{spa}, \boldsymbol{\theta}^j=\boldsymbol{\theta}, w^j=w$ \;
    Let $F^j$ be a length-\textit{N} binary vector whose $i-$th entry is 1 only if $i \in U^j$
}
Denote $J_0$ as the one such that $F_{J_0}$ is closet to the mean of $\{F_j, j=1,...,J\}$
\end{algorithm*}

\section{Experiments on synthetic data}
We evaluated the performance of SRMR and selected baseline methods on a comprehensive setup of synthetic datasets, and evaluated the overall performance  of SRMR in solving the spatially dependent mixture regression problem with different number of mixture models, level of linear dependency, spatial distribution, and ratio of outliers.

\subsection{Baseline Methods}

We collected in total nine existing methods to represent the current works. In the field of mixture regression, Pan et al. \cite{wu2016new} proposed DC-ADMM which cluster mixture content in a group pursuit way. It has an implementation as ``PRclust'' \footnote{https://github.com/ChongWu-Biostat/prclust} \textit{R} package. In the field of robust mixture regression, we collected two state-of-the-art algorithms, Trimmed Likelihood Estimation (TLE) and Component-wise adaptive Trimming Likelihood Estimation (CTLE) from \textit{R} package ``RobMixReg" \footnote{https://cran.r-project.org/web/packages/RobMixReg/} in CRAN \cite{chang2020robmixreg}. In the field of spatial smooth regression, we collected four algorithms, spatially clustered coefficient regression (SCC)\footnote{https://github.com/furong-tamu/Supplementary-files-for-SCC} \cite{li2019spatial}, Spatialculster \footnote{https://github.com/mpadge/spatialcluster} \cite{guo2008regionalization}, Spdep \footnote{https://github.com/r-spatial/spdep/} \cite{bivand2011spdep}, and ClustGeo \footnote{https://cran.r-project.org/web/packages/ClustGeo/} \cite{chavent2018clustgeo}. However, only ClustGeo can be executed under our formulation. In the field of segmentation methods based on Markov Random Field, we collected two methods FRGMM \footnote{https://sites.google.com/site/nguyen1j/home/10-code} \cite{nguyen2012fast} and mrf2d \footnote{https://freguglia.github.io/mrf2d/} \cite{freguglia2020mrf2d}. However, these two methods aim to clustering image pixels, which requires natural spatial orders from neighborhood pixels as input, and hence cannot be applied to solve our problem. Finally, we used four baseline methods DC-ADMM, TLE, CTLE, and ClustGeo to perform comparison experiments. All baseline methods used with their default parameters, except \textbf{nit} parameter in TLE and CTLERob were set as 10. For DC-ADMM, we used \textbf{stability-prclust} function to select the best parameter, followed by the instruction. For ClustGeo, we used \textbf{choicealpha} function to select best parameter.

\subsection{Experimental setup}
To simulate spatially dependent linear relationships, we first generate a univariate independent variable $x$ from uniform distribution $X \sim U(-2,2)$ and dependent variable $y$ by $y_i=\beta_k x_i + \sigma_k,k=1,...,K,i=1,...,n$, where $K$ is number of mixture models and $\beta$ is regression coefficient. Spatial coordinate of each sample $s_i$ was generated from a multivariate normal distribution $\mathcal{N}(\mu, \Sigma)$, where $\mu$ determines the center and $\Sigma$ determines the range and shape of each spatial region. We use $\mu_1=[1,1]^T, \mu_2=[-1,-1]^T, \Sigma=diag(0.1,0.1)$ as the default experimental setting, i.e. $K=2$ of two distinct and non-uniformly distributed spatial regions. The two types of outliers wer further simulated. We simulated Type 1 outliers by a rejection sampling approach. Specifically, we first samples independent $(x,y)$ from $(U(-2,2),U(-8,8))$ and only accept the ones whose Euclidean distance to the regression lines larger than two as Type 1 outliers. To simulate the Type 2 outliers of a certain ratio, we randomly select the ratio of samples and reverse their spatial coordinate $s_i=(c^1_i,c^2_i)$ by $s_i^o=(-c^1_i,-c^2_i)$. 

We conducted the synthetic data based experiments for three types of method evaluation:\\
\textbf{(1)} We evaluated the general performance of SRMR and baseline methods in solving the spatially dependent mixture regression problem by the following experimental setups (Fig 1A). Each time we perturbed one of the five factors and fixed the others, including number of mixture regression models $K=\{2,3,4\}$, total sample size $N=\{100,200,400\}$, error of linear regression $\sigma=\{0.1,0.2,0.5\}$, rate of samples belong to (model$_1$, model$_2$, outliers)=$\{(0.4,0.4,0.2),(0.5,0.3,0.2),(0.6,0.2,0.2)\}$ (only for $K$=2), and coefficients of linear regression model $\boldsymbol{\beta}=\{ (1.5,1.0),(1.5,0.1),(1.5,-1.2)\}$  (only for $K$=2).\\ 
\textbf{(2)} We validated the robustness of SRMR and baseline methods in handling the two types of outliers, namely Type 1 and 2 outliers by perturbing their ratio from 10$\%$ to 20$\%$ (Fig 1B). \\
\textbf{(3)} We validated the capability of SRMR and baseline methods in detecting different shapes and distributions of spatial regions. We simulated the spatial coordinates from a multivariate normal distribution or a multivariate uniform distribution, the former one simulates a round and dense spatial region while the later one generates uniformly distributed 2D coordinates. The simulated shapes are showcased in Fig 1C. In addition, we also evaluated if SRMR is sensitive to different relative positions of the spatial regions. We simulated two types of relative location of spatial regions, namely (i) diagonal distribution by setting $\mu_1=[1,1]^T,\mu_2=[-1,-1]^T$ and (ii)  horizontal distribution by setting $\mu_1=[0.5,0]^T,\mu_2=[-0.5,0]^T$. To simulate spatial regions of imbalanced densities, we perturbed the covariance matrix of the spatial coordinates from $diag(0.1, 0.1)$ to $diag(0.5, 0.1)$.

In summary, we set ten perturbation scenarios (Fig 1), each contains 2-3 different parameter settings. We conducted 100 replicates for each parameter set in each scenario. In total, we obtained 2,500 synthetic data sets. The mean value of evaluation metrics were used for performance evaluation.

\begin{figure*}[!] 
    \centering
    \includegraphics[width=17cm]{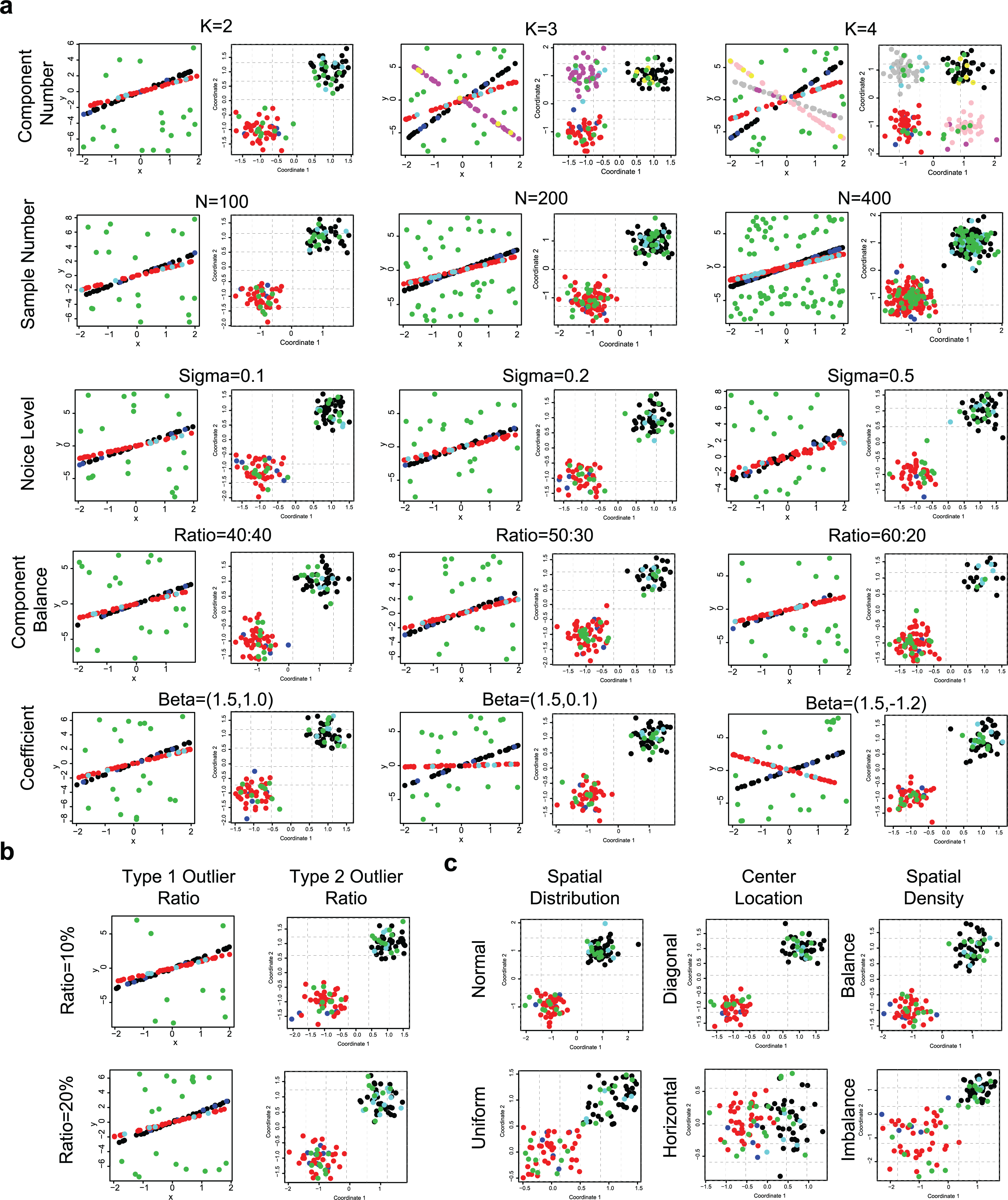}
    \caption{Experiment Setting. Sub-figures without grid represent linear relationship and sub-figures with grid represent spatial coordinates. For (b) and (c), we only show partial plot which control factor is changed instead of full plot (linear relationship and spatial coordinate) as (a). (a) contains five different scenarios in terms of mixture regression. (b) contains two scenarios to deal with Type 1 and Type 2 outliers. (c) contains three scenarios for detecting different shapes and distributions of spatial regions. }
\end{figure*}

\subsection{Evaluation Metrics}
We evaluated the performance of SRMR and baseline methods on synthetic datasets, based on how accurate the methods can identify the simulated mixture regression models and corresponding spatial regions, and distinguish the two types of outliers. Four evaluation metrics were utilized in the synthetic data based evaluations:   

1) \textit{Rand Index (RI)}: $= \frac{\text{number of agreeing pairs}}{\text{number of total pairs}}$ computes a similarity measure between two clusters by considering counting the sample pairs that are assigned in the same or different clusters in the predicted and true clusters.

2) \textit{Adjust Rand Index (ARI)}: $= \frac{\text{RI- Expected (RI)}}{\text{max(RI) - Expected(RI)}}$, which is a corrected-for-chance version of RI.

3) \textit{Accuracy Rate (ACC)} for outliear detection. ACC: $= \frac{\text{detected true outliers}}{\text{true outliers}}$ measures the accuracy for distinguishing the Type 1 and Type 2 outliers.

4) \textit{Error of Predicted Coefficients (PCE)}: $ = \sum_{k=1}^{K} (\beta_k-\beta^p_{l(k)})^2 $ measures the distance between the true regression coefficient $\beta_k$ of the regression components $k=1,...,K$ and predicted regression coefficient $\beta^p$. Here $l(k)=\underset{j}{\operatorname{argmin}} (\beta_k-\beta^p_j)^2$, i.e., $\beta^p_l(k)$ is the predicted coefficient closest to $\beta_k$.


\subsection{Performance}
We organized the synthetic data experiment results in Table 1 into three sections: mixture regression, robustness and spatial patterns. Overall, SRMR outperforms baseline methods in all 10 experiment settings under almost all evaluation metrics. 

In Table 1, the first section (1st- 5th blocks) illustrated the performance of SRMR and other methods in terms of the accuracies in detecting the heterogeneous linear dependencies in different scenarios, with regards to sample size, number of components, noise level, cluster balance and strength of regression coefficients. SRMR could detect the clusters and regression coefficients for each cluster very accurately, for different sample sizes, components, and it is robust to the different noise levels, imbalance of cluster sizes and small regression coefficients. Notably, because it incorporates spatial information, it is able to differentiate two clusters with very similar regression coefficients but different spatial locations. Since DC-ADMM and ClustGeo are designed for clustering, but not regression, the evaluation metrics ACC and PCE for these two methods are filled with NaN. Although DC-ADMM proposed using a novel formulation for clustering, it cannot handle outliers or incorporate spatial information. Thus, the performance of DC-ADMM is the lowest in most of cases. As noise level of regression line increased, the power of ordinary robust mixture regression methods TLE and CTLE decreased, leading to lower RI and ARI score. When the clusters become more and more imbalanced, the RI and ARI scores of all of the baseline methods get much worse. When two clusters have very similar regression parameters, but are far away in terms of spatial locations, TLE and CTLE cannot differentiate the two clusters, as they didn't account for spatial proximity, causing low RI and ARI score.

The second section (6th-7th blocks) of Table 1 illustrated the performance of all methods in terms of robustness to outlier contamination, including Type 1 outliers and Type 2 outliers. SRMR is highly robust to both regression outliers and spatial outliers, and the clustering accuracies and parameter estimates are almost unaffected in the presence of outliers. This is because SRMR adopted a trimmed likelihood approach, and it is expected that the outliers will not be taken into model estimations. Since DC-ADMM and ClustGeo are not designed to handle the neither Type 1 or Type 2 outliers, their performance consistently worse than TLE, CTLE, and SRMR. While TLE and CTLE could handle regression outliers, they have no control over the spatial proximity, and hence they are very sensitive Type 2 outliers, i.e., spatial outliers. ACC of TLE and CTLE is around 70\% due to spatial heterogeneity while SRMR has 100\% accuracy rate in all scenarios.

The third section (8th-10blocks) in Table 1 illustrate the performance of all methods for different spatial patterns, regarding the shape, center and density of the spatial clusters. SRMR is designed to detect heterogeneious linear dependencies that is robust to both regression outliers and spatial outliers, and its performance is consistently desirably with regards to different spatial patterns. When the spatial distribution of the clusters are changed from multivariate normal to multivariate uniform, it means the shape of the clusters are less sphear, and more diffused. When the center of spatial coordinate changed from diagonal to horizontal, the boundary of two spatial centers became blurred, meaning there are more overlap between neighbouring clusters. The performance of TLE, CTLE and ClustGeo got worse with more cluster overlaps,  while SRMR is robust to this complex situation thanks to the integration of both regression and spatial similarity. ClustGEO is sensitive to the imbalanced density of different clusters, while SRMR is unaffected. 

In summary, SRMR is the only method that could model the linear dependency between response and predictors that vary in the spatial domain, and detect clusters of observations with both similarities in regression parameters and spatial proximity. And it is robust to both outliers in regression fitting and spatial locations. It has produced highly favorable performance in different simulation settings, with regards to different levels of regression/spatial noise, outliers, and mixture imbalance.

\begin{figure*}[!] 
    \centering
    \includegraphics[width=17cm]{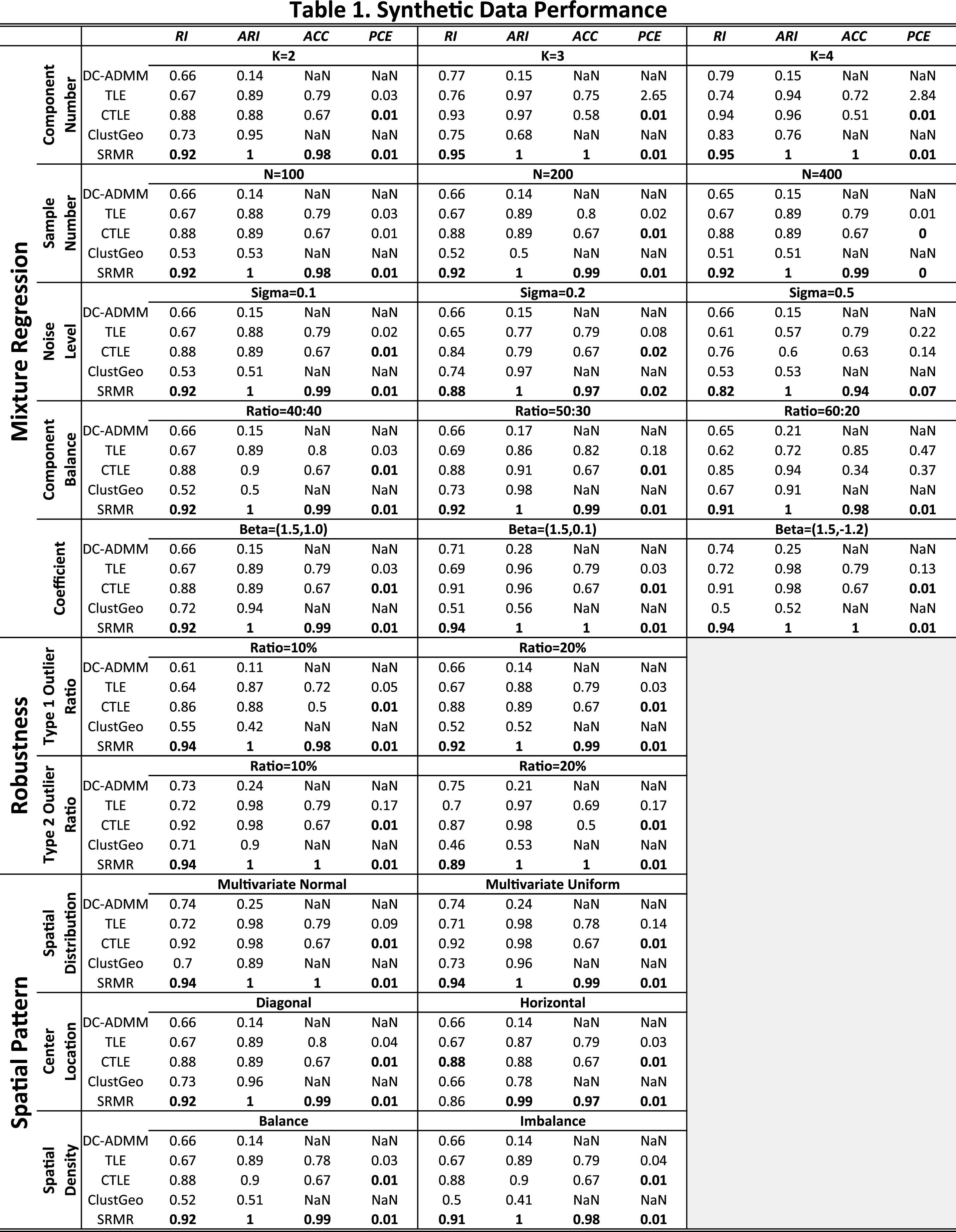}
\end{figure*}

\section{Experiments on real-world data}
We further validated SRMR on two real-world datasets, namely (1) a geospatial economics data collected from 298 cities of China and (2) a spatial transcriptomics data collected from 3,798 spatial spots on a 2D breast cancer tissue. The synthetic data based experiments clearly suggested that SRMR is the only method that can effectively solve the spatially dependent mixture regression problem compared to the baseline methods. In the real-world data based experiments, we mainly focused on illustrating the contextual meaning of the spatial regions and corresponding regression models identified by SRMR. We also evaluated the goodness of fitting and significance of the spatially dependent mixture regression models as well as the running time of the tested methods.

\begin{figure*}[!] 
    \centering
    \includegraphics[width=16cm]{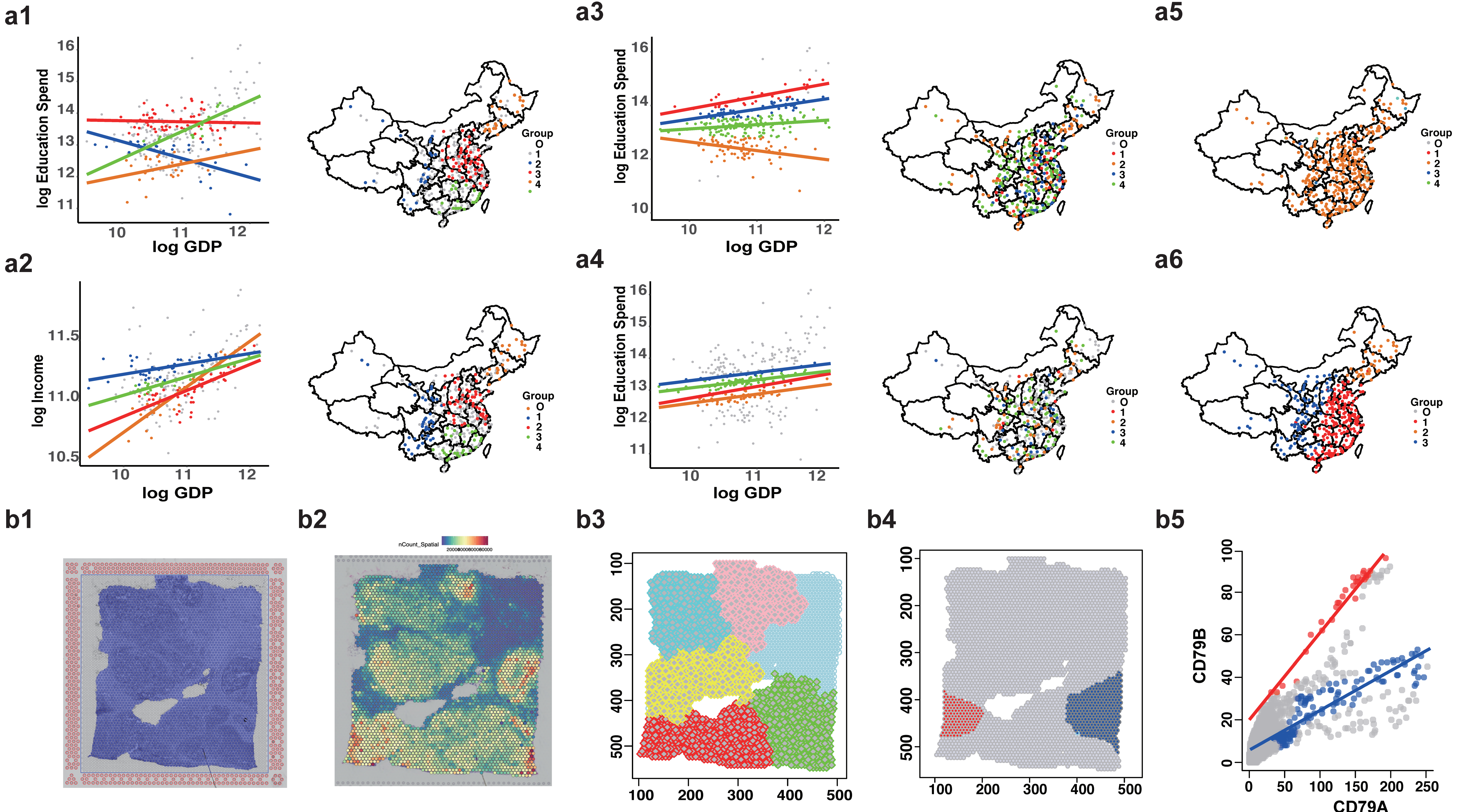}
    \caption{Real-world data based experiments. a1: SRMR, a2: SRMR, a3: TLE, a4: CTLE, a5: DC-ADMM, a6: ClustGeo; a1,a3-a6: $ES\sim GDP$, a2: $Income\sim GDP$. a1-a6: cities of different regression components are red, blue, green and orange colored, while the outliers are colored by grey. }
\end{figure*}

\subsection{Application on Geospatial Economics Data}
We collected 7 economic features, namely total GDP, public income, public spend, educational spend, technology spend, population, and averaged personal income, and latitude and longitudes, for 298 cities in China. We evaluated SRMR and baseline methods to this data set. We utilized each of the eight features as a dependent variable and selected others as independent variables When applying SRMR and other regression models, while all the features were utilized as the input of ClustGeo. Similar to the synthetic data based experiments, SRMR is the only method can identify spatially dependent mixture regression models. In contrast, TLE, CTLERob, and DC-ADMM only detected spatial independent regression models, and ClustGeo output a spatial segmentation based on all features.

For a clear visualization and explanation, we illustrated two univariate regressions of $Educational\ Spend\ (ES)\ \sim\ GDP$ and $Income\ \sim\ GDP$. For both $ES$ and $Income$, SRMR identified four spatial regions corresponding to the north-east, middle-east, south-east and west regions of China (Fig 2a1,2a2). The spatial regions detected by SRMR show distinct different dependency of $ES$ and $Income$ with $GDP$. Specifically, $ES$ is positively associated with $GDP$ in the middle-east ($ES=0.24\cdot GDP+10.9$) and north-east China ($ES=0.4\cdot GDP+9.17$). The south-east cities have more stable $ES$, which less depends on $GDP$ ($ES=0.8\cdot GDP+4.19$), while a negative association of $ES$ and $GDP$ are observed in the west cities ($ES=-0.39\cdot GDP+17.09$). The high dependency in middle-east and north-east cities and less dependency in south-east cities are consistent to our knowledge, as the middle-east and north-east China are promoting the education system  basis while the education systems south-east China are relatively stable. We also checked the cities in the west China that have high $GDP$ but low $ES$. Such cities include Dongying, Ordos, Karamay, etc., which are developing more neo energy business rather than education in the recent years. Similar observations were also made in the $Income\ \sim\ GDP$ model (Fig 2a2). The SRMR outputs suggested the personal $Income$ in the north-east, south-east and west cities less depends on $GDP$ while more positive dependency between $Income$ and $GDP$ was observed in middle-east cities, especially the well developed cities Beijing, Shanghai, Tianjin, Hangzhou, etc. On the other hand, on both $Educational\ Spend\ (ES)\ \sim\ GDP$ and $Income\ \sim\ GDP$, TLE and CTLERob failed to identify such spatial dependent and contextual meaningful patterns while both of them tend to over-fit the mixture of regressions (Fig 2a3, 2a4). DC-ADMM identified all cities as one class (Fig 2a5) while ClustGeo identified three distinct non-overlapping spatial regions without offering explainable regional specific feature dependencies (Fig 2a6). 

\subsection{Application on Spatial Transcriptomics Data}
10x Genomics spatial transcriptomics (ST) is a recent commercialized technique to measure spatial coordinates associated gene expression signal from a biological tissue sample, which it has a broad utilization in biomedical research. A typical ST data is a matrix consisting of $\sim$15,000 genes (rows) in $\sim$4,000 individual spatial spots (columns), and each spot has a 2D spatial coordinate (Fig 2b1). The spatial spots are uniformly distributed. A key challenge in ST data analysis is to infer the spatially dependent and biologically meaningful functional variations from the high dimensional feature matrix (genes by spatial spots). Here we illustrate that SRMR enables a new type of ST data analysis by simultaneously identify spatial regions in which the expression level of genes show different level of dependency, which directly annotate the biological meaning of each detected region.

We applied SRMR and baseline methods on the v1.1 ST data of breast cancer provided by 10xgenomics.com, consisting of 13,161 genes and 3,798 spatial spots.  We first selected 500 genes that having high expression level and having known tumor micro-environment related functions. We fit the regression model Gene$_1\ \sim$ Gene$_2$ for each pair of the 500 genes by using SRMR, TLE, CTLERob and DC-ADMM and conducted ClustGeo by using all the 500 genes. Similar to the synthetic and Geospatial data, SRMR is the only method that detected spatially dependent mixture regression models in the ST data. General spatial segmentation, such as ClustGeo, identifies spatial regions by using the whole feature matrix (Fig 2b3), which is consistent to the distribution of the averaged gene expression signal level (Fig 2b2). On the other hand, we identified more than 500 overlapped spatial regions by using SRMR, each having varied dependency among certain genes. Fig 2b4 showcased two distinct spatial regions only identified by SRMR, which have varied dependency between the CD79A and CD79B genes as shown in Fig 2b5. CD79A/B are key genes involved in maturation and functional variation of B cells. The varied dependency of CD79A and CD79B characterizes distinct sub-regions in one breast cancer tissue that potentially have different immune activities and responses to immuno-therapy.

In summary, compared with baseline methods, SRMR is the only method can effectively solve the spatially dependent mixture regression problem on the two real-world data. For the analysis of a single regression model in the real-world data, the running time of SRMR, TLE and CTLERob are about 15s, 10s and 2s, respectively. The running time of SRMR is slower, but also comparable to the baseline robust mixture regression approaches. The running time of DC-ADMM and ClustGeo are about 0.01s.

\section{Conclusion}
We developed a new statistical model of high dimensional data with matched spatial information, namely spatially dependent mixture regression. We also developed spatial robust mixture regression (SRMR) analysis as an effective solution of the problem. SRMR is empowered by an inference scheme to assess statistical significance of spatial dependent finite mixture regression models. On both synthetic and real-world data based experiments, we demonstrated that SRMR is the only capability can solve the spatially dependent mixture regression problem. Particularly, SRMR enables a new type of spatial segmentation analysis by detecting large sets of spatial regions having varied dependency among certain features. Compared with conventional spatial segmentation analysis, the regions identified by SRMR characterize more spatial dependent variations conceived in the data and enable better contextual explanation. The source codes of SRMR and the analysis of this study are provided at https://github.com/changwn/SRMR.

\bibliographystyle{IEEEtranN}
{\scriptsize
\bibliography{ref.bib}}

\end{document}